\documentclass[journal]{IEEEtran}
\usepackage{lineno}
\usepackage{epsfig}
\usepackage{chngpage}
\usepackage{float}
\usepackage{amsmath}
\usepackage{graphicx}
\usepackage{array}
\usepackage{diagbox}
\usepackage{epstopdf}
\usepackage{algorithm}
\usepackage{algorithmicx}
\usepackage{algpseudocode}
\usepackage{threeparttable}
\usepackage{subfigure}
\usepackage{multirow}
\usepackage{longtable}
\usepackage[table,xcdraw]{xcolor}
\usepackage{url}
\usepackage{booktabs}
\usepackage{multirow}
\usepackage[colorlinks,linkcolor=blue]{hyperref}
\usepackage{algorithm}
\usepackage{algorithmicx}
\usepackage{algpseudocode}
\usepackage{amsmath}
\usepackage{amssymb}
\begin{document}
%
\title{Random Walks: A Review of Algorithms and Applications}

\author{Feng Xia,~\IEEEmembership{Senior Member,~IEEE,}
        Jiaying Liu,
        Hansong Nie,
        Yonghao Fu,
        Liangtian Wan,~\IEEEmembership{Member,~IEEE,}
        Xiangjie Kong,~\IEEEmembership{Senior Member,~IEEE}
\thanks{This work was partially supported by the National Natural Science Foundation of China (61872054), and the Fundamental Research Funds for the Central Universities under Grant (DUT19LAB23, DUT18JC09).}
\thanks{The authors are with the Key Laboratory for Ubiquitous Network and Service Software of Liaoning Province, School of Software, Dalian University of Technology, Dalian 116620, China.}
\thanks{F. Xia is also with School of Science, Engineering and Information Technology, Federation University Australia, Ballarat, VIC 3353, Australia}
\thanks{Corresponding author: Xiangjie Kong; email: xjkong@ieee.org.}
}

\markboth{IEEE Transactions on Emerging Topics in Computational Intelligence,~Vol.~0, No.~0, May~2019}%
{Shell \MakeLowercase{\textit{ Xia et al.}}: Random Walks: A Review of Algorithms and Applications}
%



\maketitle

\begin{abstract}
A random walk is known as a random process which describes a path including a succession of random steps in the mathematical space. It has increasingly been popular in various disciplines such as mathematics and computer science. Furthermore, in quantum mechanics, quantum walks can be regarded as quantum analogues of classical random walks. Classical random walks and quantum walks can be used to calculate the proximity between nodes and extract the topology in the network. Various random walk related models can be applied in different fields, which is of great significance to downstream tasks such as link prediction, recommendation, computer vision, semi-supervised learning, and network embedding. In this paper, we aim to provide a comprehensive review of classical random walks and quantum walks. We first review the knowledge of classical random walks and quantum walks, including basic concepts and some typical algorithms. We also compare the algorithms based on quantum walks and classical random walks from the perspective of time complexity. Then we introduce their applications in the field of computer science. Finally we discuss the open issues from the perspectives of efficiency, main-memory volume, and computing time of existing algorithms. This study aims to contribute to this growing area of research by exploring random walks and quantum walks together.
\end{abstract}


%
\IEEEpeerreviewmaketitle

\section{Introduction}
A random walk is a random process in the mathematical space. It describes a path consisting of a succession of random steps in the mathematical space. It is firstly introduced by Pearson in 1905~\cite{pearson1905problem}. Spitzer~\cite{spitzer2013principles} gives a complete review of random walks for mathematical researchers and clearly presents the mathematical principles of random walks. Random walks can be used to analyze and simulate the randomness of objects and calculate the correlation among objects, which are useful in solving practical problems. It is fast becoming a key instrument in the fields of computer science, physics, chemistry, biology, economics, etc.

In the mathematical space, a simple random walk model is a random walk on a regular lattice, in which one point can jump to another position at each step according to a certain probability distribution. When it is applied on a specific network, the transition probability between nodes is positively relevant to their correlation strength. That is, the stronger their association is, the greater the transition probability is. After enough steps, we can obtain a random path that can describe the network structure.

The most typical random walk based algorithms in computer science area is PageRank~\cite{page1999pagerank}. It calculates the importance of web pages by walking randomly among them. Researchers have developed a series variants of Random Walk, such as personalized PageRank~\cite{fogaras2005towards,haveliwala2003topic}, random walk with restart (RWR)~\cite{pan2004automatic}, and lazy random walk (LRW)~\cite{shen2014lazy}.

Quantum walks are first proposed by Aharonov et al.~\cite{aharonov1993quantum} in 1993. Quantum walks can be regarded as the counter part of classical random walks in quantum mechanics. The main difference between classical random walks and quantum walks is that quantum walks don't converge to some limiting distributions. They can spread significantly faster or slower than classical random walks because of quantum interference. Compared to classical random walk based algorithms, quantum walk based algorithms have lower time complexity~\cite{childs2003exponential,farhi1998quantum,grover1996fast,shenvi2003quantum}. They can provide an exponential speedup over any classical algorithm~\cite{childs2003exponential}.
Quantum walk based algorithms can be roughly divided into two categories: discrete time based algorithms and continuous time based algorithms~\cite{kempe2003quantum}.

A random walk is implemented by utilizing the network topology, so it can also be used to calculate the proximity between nodes. For example, researchers have introduced algorithms based on random walks in the area of collaborative filtering~\cite{adomavicius2004recommendation,brand2005random,fouss2005novel,fouss2007random,yildirim2008random,jamali2009trustwalker}. Compared with other alternative approaches, random walk based algorithms can incorporate a great deal of contextual information. As same as collaborative filtering, link prediction and recommender system also aim to calculate the $k$-most-close nodes for the selected node. Hence, random walks are also effective in link prediction and recommendation system~\cite{gori2007itemrank,gori2006research,xia2016scientific,xia2014mvcwalker,sarkar2012tractable,liu2010link,backstrom2011supervised,liu2017inferring}.
Random walks can also be applied in
computer vision~\cite{shen2014lazy,meila2001random,gorelick2006shape,grady2006random,grady2004multi,grady2005multilabel,qiu2005image,qiu2006robust,dong2016sub,li2016visual}, semi-supervised learning~\cite{zhu2006semi,zhu2003semi,szummer2002partially,tishby2001data,azran2007rendezvous},
network embedding\cite{perozzi2014deepwalk,grover2016node2vec}, and complex social network analysis~\cite{sarkar2011random}. There are also some literature illustrating the applications of random walks on graphs~\cite{lovasz1993random,aldous2002reversible}, text analysis~\cite{amancio2015concentric}, science of science~\cite{amancio2015comparing}, and knowledge discovery~\cite{de2017knowledge}. Quantum walks are often used to accelerate classical algorithms. It can be used to decision trees~\cite{farhi1998quantum}, search problems~\cite{grover1996fast,shenvi2003quantum}, and element distinctness~\cite{buhrman2001quantum,ambainis2007quantum}.

In this paper, we provide a comprehensive review of random walks. To the best of our knowledge, it is the first time to review classical random walks and quantum walks together. We summarize random walks in the field of computer science from the perspectives of basic concept, algorithm and application. We also compare these algorithms systematically. In addition, some open issues of random walks and quantum walks are presented.

\begin{figure*}
	\centering
	\includegraphics[width=18cm]{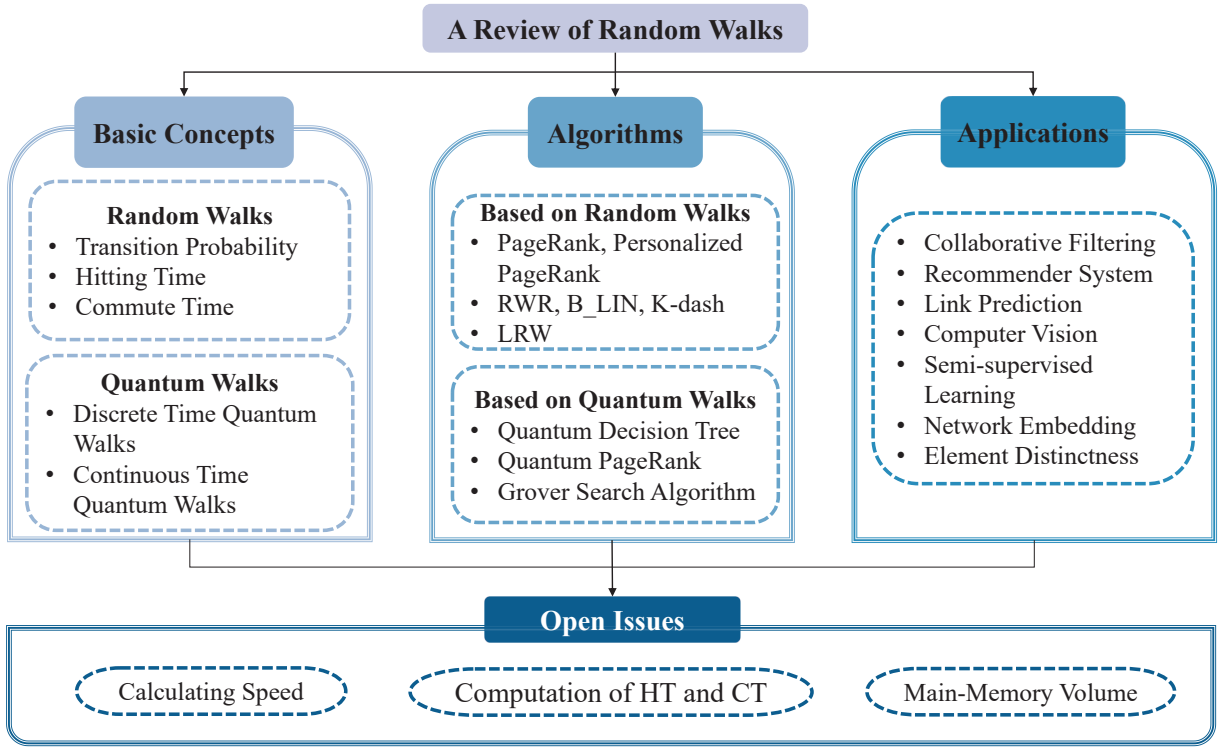}\\
	\caption{The framework of random walks review.}
	\label{structure}
\end{figure*}

In the rest of the paper, we first introduce basic concepts and notations of classical random walks and quantum walks in Section~\ref{section2}. In particular, we introduce quantum walks from two aspects: discrete time quantum walks and continuous time quantum walks. In Section~\ref{section3}, we focus on illustrating some typical algorithms based on classical random walks and quantum walks. We also make an analytical comparison between these algorithms. In Section~\ref{section4}, we show the application scenarios of different algorithms and identify their advantages and disadvantages. Section~\ref{section5} highlights the problems and future directions. Finally, the work is concluded in Section~\ref{section6}. The overall structure of this paper is summarized in Fig.~\ref{structure}.

\section{Preliminaries and Notations}\label{section2}
\par In this section, we will introduce the basic concepts and notations about random walks, including classical random walks and quantum walks. TABLE~\ref{TabOfNo} lists the commonly used notations in this paper.

\begin{table}
	\caption{Description of notations.}
	\centering
	\begin{center}
		\begin{tabular}{p{2.5cm}<{\centering}|p{5cm}<{\centering}}
			\hline
			\hline
			Notation & Description \\  [1.5pt]
			\hline
			$\{ \xi_{t}, t=0,1,2,... \}$  &  the random variable in a sequence \\  [1.5pt]
			$P_0$  &  the initial distribution of a random walk  \\  [1.5pt]
			$t$  &  the step number of a random walk  \\  [1.5pt]
			$P_t$  &   the distribution after $t$ steps in a random walk  \\  [1.5pt]
			$G$  &  a connected graph  \\  [1.5pt]
			$V$  &  the set of nodes in $G$  \\  [1.5pt]
			$E$  &  the set of edges in $G$  \\  [1.5pt]
			$n$  &  the number of nodes in $G$  \\  [1.5pt]
			$A\in \mathbb{R}^{n\times n}$  &  the adjacency matrix of $G$  \\  [1.5pt]
			$p_{ij}$  &  the probability (single step) from node $i$ to node $j$  \\  [1.5pt]
			$p_{ij}^{(t)}$  &  the probability ($t$ steps) from node $i$ to node $j$  \\[1.5pt]
			$M\in \mathbb{R}^{n\times n}$  &  the matrix of transition probabilities of $G$  \\[1.5pt]
			$D\in \mathbb{R}^{n\times n}$  &  a diagonal matrix\\  [1.5pt]
			$L$  &  the Laplacian matrix of $G$ \\  [1.5pt]
			$h_{ij}$  &  the hitting time between node $i$ and node $j$  \\  [1.5pt]
			$C_{ij}$  &  the commute time between node $i$ and node $j$ \\  [1.5pt]
			$\mathcal{N}_i$  &  the neighbor set of node $i$  \\  [1.5pt]
			$\vert \psi_{x_{0}}  \rangle$ &  the wave-packet localizing around the position $x_0$ \\  [1.5pt]
			$\mathcal{P}$ & the momentum operator \\  [1.5pt]
			$U_{l}$ &  the unitary operator \\  [1.5pt]
			$S_{z}$ &  the $z$ component of the spin \\  [1.5pt]
			$\vert \uparrow \rangle$, $\vert \downarrow \rangle$ &  two eigenstates of $S_z$ \\  [1.5pt]
			$\mathcal{H}_p$  &  the Hilberet space \\  [1.5pt]
			$\mathcal{H}_c$  &  the coin-space  \\ [1.5pt]
			$\otimes$  &  tensor product \\ [1.5pt]
			$H$ &  Hadamard coin \\ [1.5pt]
			$\hat{H}$  &  an infinitesimal generator matrix and the Hamiltonian function\\  [1.5pt]
			\hline
		\end{tabular}
	\label{TabOfNo}
	\end{center}
\end{table}

\subsection{Classical Random Walks}
A random walk is known as a random process. It describes a path consisting of a succession of random steps on some mathematical space, which can be denoted as $\{ \xi_{t}, t=0,1,2,... \}$ where $\xi_{t}$ is a random variable describing the position of a random walk after $t$ steps. The sequence can also be regarded as a special category of Markov chain. In the initial state of a random walk, the position $\xi_{0}$ may be fixed or drawn from some initial distribution $P_{0}$~\cite{lovasz1993random}. We can represent the distribution of position after $t$ steps as follows:
\begin{equation}
P_{t}(i)=Pr(\xi_{t}=i)
\end{equation}
where $P_{t}(i)$ is the probability that the random walk visits the position $i$ after $t$ steps. If the walk locates at the position $i$ after $t$ steps, the single step transition probability refers to the probability that the random walk can move to the position $j$ at after the next step. It is represented as $p_{ij}$ and can be calculated as:
\begin{equation}
p_{ij}=Pr(\xi_{t+1}=j \vert \xi_{t}=i).
\end{equation}
Further, the $t$ steps transition probability is defined as follows:
\begin{equation}
p_{ij}^{(t)}=Pr(\xi_{t}=j \vert \xi_{0}=i).
\end{equation}

From the perspective of graph representation, let $G=(V,E)$ be a connected graph, where $V$ is the vertex set and $E$ is the edge set. The adjacency matrix of $G$ is denoted as $A\in \mathbb{R}^{n\times n}$, where $n$ is the number of nodes in $G$. $A_{ij}$ denotes the weight of edge from the node $i$ to the node $j$. Then the transition probability (single step) from node $i$ to node $j$ on the graph can be defined as:
\begin{equation}
p_{ij}=\frac{A_{ij}}{\sum_{j\in V}A_{ij}}.
 \end{equation}
Further, let $M=(p_{ij})_{i,j\in V}$ be the matrix of transition probabilities on $G$. Then we can define $D$ which denotes a diagonal matrix as:
\begin{equation}
D_{ii}=1/\sum_{j\in V}A_{ij}.
\end{equation}
Thus we can redefine the transition probability matrix $M$ of graph $G$ as:
\begin{equation}
M=DA.
\end{equation}
The rule of a random walk can be expressed as:
\begin{equation}
P_{t+1}=M^{T}P_{t}
\end{equation}
where $P_{t}$ can be viewed as a vector in $\mathbb{R}^{\lvert V\rvert}$. Its $i$-th element means the probability that the random walk from the initial node $v_0$ reaches the $i$-th node after $t$ steps. We can calculate $P_{t}$ as:
\begin{equation} \label{tp}
P_{t}=(M^{T})^{t}P_{0}.
\end{equation}
The Laplacian matrix of $G$ can be defined as follows:
\begin{equation}\label{lp}
L = D-A.
\end{equation}

\par \textbf{Hitting time.}
Hitting time $h_{ij}$ can be considered as the expected number of steps before node $j$ visited in a random walk starting from node $i$~\cite{lovasz1993random}. The recursive definition of hitting time is as follows:
\begin{equation}
 h_{ij} =
\begin{cases}
1+\sum_{k \in \mathcal{N}_i} p_{ik}h_{kj} &$if$ \ i \neq j,\\
0 &$if$ \ i = j
\end{cases}
\end{equation}
where $h_{kj}$ denotes the hitting time from node $k$ to node $j$ and node $k$ is a direct neighbor of node $i$. $p_{ik}$ is the transition probability from node $i$ to node $k$. $\mathcal{N}_i$ is the neighbor set of node $i$~\cite{sarkar2011random}.

The hitting time matrix is not symmetric even in a regular graph. Another important fact about hitting time is proved by Lovasz~\cite{lovasz1993random}: hitting time follows the triangle inequality.

\par
\textbf{Commute time.} Commute time from node $i$ to node $j$ is defined as:
\begin{equation}
C_{ij} =h_{ij} + h_{ji}
\end{equation}	
which means the excepted number of steps in a random walk starting at $i$, before accessing the node $j$ and then reaching the node $i$ again~\cite{lovasz1993random}.	
In order to research the commute time on undirected graphs, Chandra et al.~\cite{doyle1984random} give an electrical network view. They compare the commute time between two nodes on a graph to the resistance on an electrical network. They give some intuitions about commute time on undirected graphs:
\begin{itemize}
	\item The smaller resistance can make the current go through more easily on electrical networks. The shorter commute time can make random walkers diffuse easier on undirected graphs.
	\item Commute time should be robust to small perturbation so that removing or adding a few resistances do not change much on an electrical network.
\end{itemize}

\subsection{Quantum Views of Random Walks}
\par
The scalable quantum computer is a topical issue so that approaches of quantum computation are popular topics nowadays. Quantum walks are the corresponding part of classical random walks in quantum mechanics. The main difference between them is that quantum walks don't converge to some limiting distributions. Due to the quantum interference, quantum walks can spread significantly faster or slower than classical random walks. Existing literature gives us explicit introduction to quantum random walk in a comprehensive way~\cite{aharonov1993quantum,kempe2003quantum,venegas2012quantum,brun2003quantum}.

In quantum mechanics, let $\vert \psi_{x_{0}}  \rangle $ denote a wave-packet which localizes around a position $x_0$. $\mathcal{P}$ is a momentum operator. The translation with length $l$ of a particle can be expressed as the unitary operator $U_l$, which can be calculated as~\cite{kempe2003quantum}:
\begin{equation}
U_l=exp(-i \mathcal{P} \hat{h})
\end{equation}
where $\hat{h}$ is reduced Planck constant which is the smallest unit of to measure angular momentum. Meanwhile, it satisfies the following formula:
\begin{equation}
U_l \psi_{x_{0}}=\psi_{x_{0}-l}
\end{equation}
where we can set $\hat{h}=1$ to simplify the notation.

We can assume that the particle has a spin-$1/2$ degree of freedom and represent the operator corresponding to the $z$ component of the spin as $S_z$. The eigenstates of $S_z$ are $\vert \uparrow \rangle$ and $\vert \downarrow \rangle$. A spin-$1/2$ particle can be described by a 2-vector:
\begin{equation}
	\vert \Psi \rangle =( \vert \widetilde \psi ^{\uparrow} \rangle, \vert \widetilde \psi ^{\downarrow} \rangle )^T
\end{equation}
where $\vert \widetilde \psi ^{\uparrow} \rangle$ is the component of the wave-function of the particle in the spin-$\vert \uparrow \rangle$ space. $\vert \widetilde \psi ^{\downarrow} \rangle$ is the component of the wave-function of the particle in the spin-$\vert \downarrow \rangle$ space.

The concept of quantum walks is firstly proposed by Aharonov et al.~\cite{aharonov1993quantum} in 1993. Kempe~\cite{kempe2003quantum} presents two kinds of quantum walk including discrete time quantum walks and continuous time quantum walks. We will introduce an easy example in one-dimensional space to help readers quickly understand the basic ideas of discrete time quantum walks and continuous time quantum walks.

\subsubsection{Discrete Time Quantum Walks}
We can define a space $\mathcal{H}=\mathcal{H}_{p} \otimes \mathcal{H}_{c}$ for one dimensional quantum walks~\cite{kempe2003quantum}. $\mathcal{H}_{p}$ denotes the Hilbert space which is spanned by the positions of the particle. For one dimensional Hilbert space, it can be represented as:
\begin{equation}
\mathcal{H}_{p}=\lbrace \vert i\rangle : i \in Z\rbrace
\end{equation}
where $\vert i \rangle$ a particle localized at the position $i$.
${\mathcal{H}_{c}}$ denotes the coin-space which is spanned by two basic states  $\lbrace \vert \uparrow \rangle,\vert \downarrow \rangle\rbrace$. The unitary operation $S$ defines the conditional translation on space $\mathcal{H}$:
\begin{equation}
S =\vert \uparrow\rangle\langle\uparrow\vert\otimes\sum_{i}{\vert i+1\rangle\langle i \vert +\vert\downarrow\rangle\langle \downarrow\vert \otimes\sum_{i}{\vert
i-1\rangle}}\langle i \vert
\end{equation}
where $i\in Z$, $\otimes$ is the tensor product which separates two degrees of freedom, spin and space, and will allow us to view the resulting correlations between these two degrees of freedom more clearly~\cite{kempe2003quantum}.
$S$ can realize the following equations:
\begin{equation}
S(\vert \uparrow \rangle \otimes \vert i \rangle) = \vert \uparrow \rangle \otimes \vert i+1\rangle,
\end{equation}
\begin{equation}
S(\vert \downarrow\rangle\otimes \vert i\rangle) = \vert \downarrow\rangle\otimes\vert i-1\rangle.
\end{equation}
It means that the particle jumps right if it has spin up and left if it has spin down.

$C$ is a unitary transformation which can rotate the spin in $H_{c}$. One of the most frequently used unitary transformation is Hadamard coin $H$~\cite{kempe2003quantum}. Here is an example of $H$:
\begin{equation}
H=\frac{1}{\sqrt{2}}\begin{pmatrix} 1 & 1 \\ 1 & -1 \end{pmatrix}\\.
\end{equation}
The Hadamard walk on $Z$ is~\cite{kempe2003quantum}:
\begin{equation}
\begin{split}
\vert \uparrow \rangle \otimes \vert 0\rangle & \xrightarrow{H} \frac{1}{\sqrt{2}} (\vert \uparrow\rangle + \vert\downarrow\rangle)\otimes\vert0\rangle \\
 &\xrightarrow{S} \frac{1}{\sqrt{2}} (\vert \uparrow\rangle \otimes \vert 1\rangle + \vert\downarrow\rangle \otimes \vert -1\rangle)
\end{split}.
\end{equation}
Then the single step quantum walk transformation can be defined as:
\begin{equation}
 U = S\cdot(C\otimes I).
\end{equation}
A quantum walk of $t$ steps is defined as the transformation $U^t$.

\subsubsection{Continuous Time Quantum Walks}
The original purpose of continuous time quantum walks is to speed up algorithms using classical random walks. The concept of continuous time quantum walks is first presented by Farhi et al.~\cite{farhi1998quantum} in 1998. The authors exploit quantum walks in the decision tree algorithm instead of classical random walks. Different from discrete time quantum walks, continuous time quantum walks don't need a coin-space $\mathcal{H}_{c}$ and take place entirely in the Hilbert space $\mathcal{H}_{p}$~\cite{kempe2003quantum}. The idea of continuous time quantum walks is from continuous time classical random walks. Kempe~\cite{kempe2003quantum} gives another expression of the continuous time random walks as:
\begin{equation}
P(t) = exp(-\hat{H}t)P(0)
\end{equation}
which is in analogy to Equation~(\ref{tp}) and $\hat{H}$ is an infinitesimal generator matrix with similar structure to $M$.

The key idea proposed by Farhi et al.~\cite{farhi1998quantum} is that the generator matrix $\hat{H}$ will become the Hamiltonian function of the process and generate an evolution $U(t)$ as follows:
\begin{equation}\label{unitaryo}
U(t)=exp(-i\hat{H}t).
\end{equation}

The connections between discrete quantum walks and continuous quantum walks are proposed by Strauch~\cite{strauch2006connecting}. The author finds that discrete quantum walks can be transferred to the continuous quantum walks by the precise limiting procedure.

\section{Algorithms Based on Random Walks}\label{section3}
In this section, we will introduce some typical algorithms based on classical random walks and quantum walks.

\subsection{Algorithms Based on Classical Random Walks }
\subsubsection{PageRank}
PageRank is first proposed by Page et al.~\cite{page1999pagerank} in 1999.  The purpose is to rank the web page in the \emph{World Wide Web} (\emph{WWW}). The network of web page is considered as a graph where web pages are considered as nodes. If there is a web page containing a hyperlink which points to another web page, then there should be a directed edge between these two nodes. The direction of the edge is as same as the web redirection. The most simple PageRank can be described by the following mathematical equation:
\begin{equation}\label{pagerank}
R(u) = c \sum_{v\in B_{u}} \frac{R(v)}{N_{v}}
\end{equation}
where $R(u)$ is the rank of the web $u$. $B_{u}$ is the set of pages pointing to page $u$, and $c$ is a normalization parameter. Let $F(v)$ be the set of pages that $v$ points to. $N_{v}$ is the number of pages in $F(v)$.

The simple version corresponds to the standing probability distribution of a random walk on the network. When a random walk quickly converges to a limiting distribution on the set of nodes, it can be regarded as rapidly-mixing. It has been proved that a random walk can be rapidly-mixing on the graph of \emph{WWW}~\cite{page1999pagerank}. The importance of a node can be regarded as the probability that the random walker reaches the node after long enough steps. The mathematical expression is:
\begin{equation}
R_{t+1}=M^{T}R_{t}
\end{equation}
where $R$ is the vector of PageRank, and $M^{T}$ is the transition probability matrix.

To improve the convergence rate of PageRank, Kamvar et al.~\cite{kamvar2003extrapolation} present a novel algorithm called Quadratic Extrapolation for PageRank computation. It accelerates the convergence of the power method. The main strategy of the algorithm is to reduce the estimation of non-main eigenvectors periodically.

The result of PageRank is independent of the keywords searched by users. To solve this problem, Haveliwala et al.~\cite{haveliwala2003topic} present personalized PageRank:
\begin{equation}
R_{t+1}=(1-\alpha)M^{T}R_{t}+\alpha p
\end{equation}
where $\alpha$ is a decay factor, $p$ is the personalized PageRank vector which reflects the importance of each node in a graph for a specific user.

\subsubsection{Random Walk with Restart}
RWR is first proposed by Pan et al.~\cite{pan2004automatic} to calculate the affinity between node $i$ and node $j$. Considering a random walk starting from node $i$, the walker can go back to node $i$ with the probability $c$, which is the difference between RWR and classical random walks. Let $u_{i}(j)$ denote the steady-state probability that the random walker will visit node $j$. The formula is:
\begin{equation} \label{RWR}
u_{i}=(1-c)M^{T}u_{i}+c e_{i}
\end{equation}
where $u_{i}$ is the probability distribution vector of RWR starting from node $i$. $e_i$ is a vector whose entry that corresponds to node $i$ equals 1, and the remaining elements being 0.

Let $G=\{V_{1}\cup V_{2},E\}$ denote a bipartite graph, where $V_{1}=\{a_{i}|1\leq i \leq k \}$ and $V_{2}=\{t_{i}|1\leq i \leq n \}$. $k$ and $n$ are the number of nodes in $V_{1}$ and $V_{2}$, respectively. The adjacency matrix $A_B$ can be written as:
\begin{equation}
A_B=
\begin{pmatrix}
0 & A \\
A^{T} & 0
\end{pmatrix}
\end{equation}
where $A$ is $k$-by-$n$ matrix. Using the bipartite structure, Sun et al.~\cite{sun2005neighborhood} propose that $u_i$ should be calculated as:
\begin{equation}
u_i=(1-c)
\begin{pmatrix}
col\_norm(A)u_{i}(k+1:k+n) \\
col\_norm(A^T)u_{i}(1:k)
\end{pmatrix}
+c e_i
\end{equation}
where $u_{i}(1:k)$ and $u_{i}(k+1:k+n)$ are the vectors of first $k$ and last $n$ elements of $u_i$, respectively. They only perform RWR on the partition that contains the query node. In other words, they present a local estimation of RWR by applying graph partition.

RWR is time-consuming when it is applied on large graphs. To fill this gap, Tong et al.~\cite{tong2006fast} present a fast RWR by low-rank approximation. The authors first rewrite RWR as:
\begin{equation}
	\begin{split}
	u_{i}&= c \hat M^{T}u_{i}+(1-c) e_{i}\\
	&= (1-c)(I-c \hat M^{T})^{-1} e_{i}\\
	&= (1-c)Q^{-1} e_{i}\\
	\end{split}
\end{equation}
where $Q=I-c \hat M^{T}$, $ \hat M$ is the normalized weighted matrix associated with $M$. Let $\hat M = \hat M_{1} + \hat M_{2}$, where $\hat M_{1}$ is the within-partition matrix and $\hat M_{2}$ is the cross-partition matrix. Then they propose B\_LIN using SVD to calculate $u_{i}$ as:
\begin{equation*}
Q_{1}=(1-c \hat M_{1})^{-1}
\end{equation*}
\begin{equation*}
\hat M_{2}=USV
\end{equation*}
\begin{equation*}
\hat\Lambda = (S^{-1}-cV Q_{1}^{-1} U)^{-1}
\end{equation*}
\begin{equation}
	u_{i}= (1-c)(Q_{1}^{-1}e_{i}+c Q_{1}^{-1} U \hat\Lambda V Q_{1}^{-1} e_{i}).
\end{equation}

Random walks are used to calculate the proximity between nodes and the specific node. If we want to find the top-k nodes, we can follow the method proposed by Fujiwara et al.~\cite{fujiwara2012fast} called K-dash to calculate the proximity of only selected nodes to find the top-k nodes. They first obtain the following equation:
\begin{equation}
	\begin{split}
	u_{i}&= c \hat M^{T}u_{i}+(1-c) e_{i}\\
	&= c(I-(1-c) \hat M^{T})^{-1} e_{i}\\
	&= cW^{-1} e_{i}\\
	\end{split}
\end{equation}
where $W=I-(1-c) \hat M^{T}$, $ \hat M$ is the column normalized weighted matrix associated with $M$. Since they don't need to calculate the proximity of all nodes, $W$ is a sparse matrix, but $W^{-1}$ may be dense. When the graph becomes large, it requires quadratic space to hold the inverse matrix which is unrealistic. Then they decompose $W$ by LU decomposition to calculate $u_{i}$ as follows:
\begin{equation*}
	W = LU
\end{equation*}
\begin{equation}
u_{i} = cU^{-1}L^{-1} e_{i}
\end{equation}
where the matrices $L^{-1}$ and $U^{-1}$ are lower and upper triangular, respectively.

\subsubsection{Lazy Random Walk}
LRW~\cite{shen2014lazy} is used to solve image segmentation problems. It first defines a graph on a given image, where every pixel is identified uniquely by a node. The similarity between node $i$ and node $j$ is defined as:
\begin{equation}
	w_{ij}=exp(-\frac{||g_i-g_j||^2}{2\sigma ^2})
\end{equation}
where $g_i$ is the image intensity value of node $i$. $\sigma$ is the user defined parameter. The degree of each node is computed as:
\begin{equation}
	d_i=\sum_{j \in V} w_{ij}.
\end{equation}
The transition probability matrix is calculated as follows:
\begin{equation}
	P_{ij} =
	\begin{cases}
	1-\alpha &  $if$ \ i=j  \\
	\alpha \cdot w_{ij}/d_{i} &$if$ \ i \sim j,\\
	0 &$if$ \ otherwise.
	\end{cases}
\end{equation}
where $i \sim j $ means the two nodes are adjacent nodes. $\alpha$ is a control parameter in the range $(0, 1)$. The equation means that the current node $i$ in LRW will have the probability $(1-\alpha)$ to stay at node $i$ and probability $\alpha$ to walk to the adjacent node. LRW will converges to a unique stationary distribution $u$ as follows:
\begin{equation}
u_i=d_i\sum_{i=1}^{N=|V|}d_i.
\end{equation}

Considering all of the above algorithms, PageRank, personalized PageRank, RWR and LRW are time-consuming on large graphs. Quadratic Extrapolation accelerates PageRank's convergence rate very well. Due to the personalized vector in personalized PageRank, it makes more sense to different users. Personalized PageRank and RWR have similar forms. RWR on a bipartite graph~\cite{sun2005neighborhood} converges faster, but has no generality. On the contrary, B\_LIN and K-dash have fast convergence rate on any graph. K-dash calculates the
proximity more exactly than B\_LIN, because LU decomposition used in K-dash is not an approximation method like SVD used in B\_LIN. TABLE \ref{COA} shows the differences among these algorithms.

\begin{table*}[!htbp]
	\renewcommand\arraystretch{1.8}
	\centering
	\caption{Comparison analytic on the classical algorithms.}
	\label{COA}
	\begin{tabular}{|c|c|c|c|}
		\hline
		 &  Formula & Time Complexity & Time-Consuming\\
		\hline
		PageRank~\cite{page1999pagerank} &  $R_{t+1}=M^{T}R_{t}$  & &Yes \\
		\hline
		personalized PageRank~\cite{haveliwala2003topic} & $R_{t+1}=(1-\alpha)M^{T}R_{t}+\alpha p$ & &Yes \\
		\hline
		RWR~\cite{pan2004automatic} & $u_{i}=(1-c)M^{T}u_{i}+c e_{i}$ & $O(|V|^{3})$ & Yes\\
		\hline
		RWR on a bipartite graph~\cite{sun2005neighborhood} & $u_i=(1-c)
		\begin{pmatrix}
			col\_norm(A)u_{i}(k+1:k+n) \\
			col\_norm(A^T)u_{i}(1:k)
		\end{pmatrix}+c e_i$  & $O(|V|^{2})$ & No \\
		\hline
		B\_LIN~\cite{tong2006fast} & $u_{i}= (1-c)(Q_{1}^{-1}e_{i}+c Q_{1}^{-1} U \hat\Lambda V Q_{1}^{-1} e_{i})$  & $O(|V|^{2})$ & No\\
		\hline
		K-dash~\cite{fujiwara2012fast} & $u_{i} = cU^{-1}L^{-1} e_{i}$ & $O(|V|+|E|)$ & No \\
		\hline
		LRW~\cite{shen2014lazy} & $u_i=d_i\sum_{i=1}^{N=|V|}d_i$ & $O(n|V|^{2})$ & Yes \\
		\hline
	\end{tabular}
\end{table*}

\subsection{Algorithms Based on Quantum Walks}
In this section, we are going to introduce some algorithms based on two quantum walk models as mentioned above. We can find some different properties between quantum walks and classical random walks.

We will separate the algorithms into two categories depending on the model they use. The first category is based on the continuous time quantum walks, such as the quantum decision tree algorithm. The other is based on discrete time quantum walks, such as the quantum PageRank algorithm.

\subsubsection{Continuous Quantum Walk Based Algorithms}
Fahri et al.~\cite{farhi1998quantum} originally present the idea of continuous time quantum walks with the example of decision tree algorithm. They choose the approach that systematically explores the whole tree with a probabilistic rule.

The authors consider decision tree nodes as quantum states in Hilbert space. Then they constructs a Hamiltonian function $\hat{H}$ which determines the time evolution of the quantum system. With the basis of Hamiltonian function, the authors present the unitary time evolution operator shown in the Equation~(\ref{unitaryo}). They find that if classical random walk based algorithms require time polynomial in $n$ to reach level $n$, quantum walks can also realize it. Moreover, if a tree is penetrable for a classical algorithm which requires time exponential in $n$, it is proved that the problem corresponding to the decision tree is solvable with this quantum algorithm in the polynomial time.

\begin{figure}[h]
	\centering
	\includegraphics[width=0.38\textwidth]{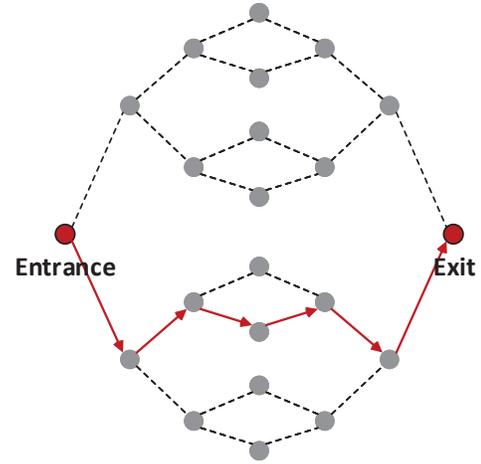}\\
	\caption{Balance tree. We want to find the node named \emph{Exit} by a classical random walk or quantum walk starting from \emph{Entrance}.}
    \label{b1}
\end{figure}

\begin{figure}[h]
	\centering
	\includegraphics[width=0.47\textwidth]{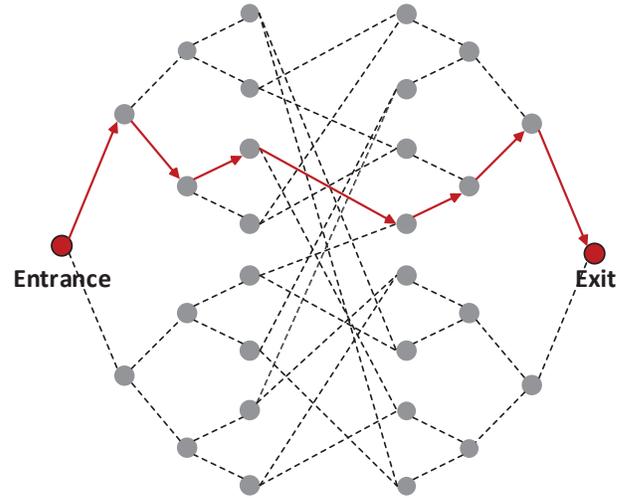}\\
	\caption{Modified balance tree. We want to find the node named \emph{Exit} by a classical random walk or quantum walk starting from \emph{Entrance}.}
    \label{b2}
\end{figure}

Childs et al.~\cite{childs2003exponential} construct an oracular problem which can be solved by a quantum walk in subexponential time. They first introduce a graph $G_r$ consisting of two balanced binary trees of height $n$ in Fig.~\ref{b1}. Then they modify the graph by randomly choosing a leaf on the left and connect it to a leaf on the right chosen at random in Fig.~\ref{b2}. Classical random walks or quantum walks go from \emph{ENTRANCE} to \emph{EXIT}.
They define a Hamiltonian function $\hat{H}$ based on $G$'s adjacency matrix to build a quantum walk on the graph.
The equation of $\hat{H}$ is:
\begin{equation}
	\langle n | \hat{H} | n' \rangle =
	\begin{cases}
	\gamma & n\neq n', nn'\in G \\
	0 & otherwise
	\end{cases}
\end{equation}
where $n$ and $n'$ are nodes of $G$. $nn'$ denotes the edge between node $n$ and node $n'$. $\gamma$ is the probability of moving to the next adjacent node. It is proved that the quantum walks are exponentially better than any classical random walks. However, this algorithm only finds the node named \emph{EXIT} without finding a path from \emph{ENTRANCE} to \emph{EXIT}.

\subsubsection{Discrete Quantum Walk Based Algorithms}
In order to study the behavior of PageRank algorithm in the quantum network, Paparo et al.~\cite{paparo2012google} present the quantum PageRank algorithm.
They give an admissible class of quantum PageRank algorithms instead of a specific definition.

The authors exploit the idea of discrete time quantum walks. They define the coin space $\mathcal{H}_{c}$ and Hilbert space $ \mathcal{H}_{p}$. The definition of coin space is similar to the one-dimension quantum walks.
\begin{equation}
\mathcal{H}_{c} = span\{\vert L\rangle ,\vert R\rangle\}.
\end{equation}

Since the PageRank algorithm is applied on the graph, the author defines the Hilbert space as the space of oriented edges:
\begin{equation}
H_{p} = span\{ \vert i \rangle_{1},\vert j\rangle_{2}\ \vert \ i,j \in N\}
\end{equation}
where $N$ denotes all the nodes on the graph. The subscripts $1$, $2$ are used to show the direction~\cite{paparo2012google}.

The authors also reveal the properties of quantum PageRank algorithm in complex real world networks~\cite{paparo2013quantum}. They find that the quantum PageRank algorithm can reveal the underlying topology of the network more univocally with respect to classical PageRank algorithms.

Considering the searching problem in the database, classical algorithms need $O(N)$ steps to find the target element, where $N$ is the number of elements. Grover~\cite{grover1996fast} proposes a novel quantum walk based algorithm to solve this problem. It is proved that the algorithm only takes $O(\sqrt{N})$ steps to find the same target. Affected by this, Shenvi et al.~\cite{shenvi2003quantum} propose a discrete quantum walk based algorithm. It can be regarded as a discrete walk on the hypercube and also achieve $O(\sqrt{N})$ searching time.

Since these algorithms are applied in different scenarios, it is hard to evaluate their performance. Compared quantum walk based algorithms with classical random walk based algorithms, computational complexity and convergence speed have been greatly improved in the quantum walks. The scope of the application is also more extensive for quantum walks. In addition, quantum walk based algorithms are better than classical random walk based algorithms in preserving the topology of network. Although the researches on quantum walks have increased in recent years~\cite{xiao2017observation,cedzich2019quantum,wang2019simulating,hou2018deterministic}, quantum walks from the perspectives of principle, mechanism and application are still worth exploring.

\section{Applications of Random Walks}\label{section4}
Random walks have been successfully applied in various areas of computer science such as recommender system, computer vision, and network embedding. In this section, we select some major applications to illustrate the effectiveness and practicability of random walks in this section.

\subsection{Collaborative Filtering}
Collaborative filtering is a method of making automatic predictions about the interests of a user by collecting preferences from many users. It assumes that two people who have the same taste on one issue will have the same interest on the other issues.

Much literature has recorded methods of collaborative filtering with successful demonstrations of Bayesian, nonparametric, linear methods, etc. All these methods are essentially the same. They all match the individual to others based on their choices and use combination of their experiences to predict future choices.

Brand et al.~\cite{brand2005random} introduce a random walk view to collaborative filtering. They want to study affinity relations on the association graph of a relational database to find out what products a customer wants to buy next.

\begin{figure}[htbp]
  \centering
  \includegraphics[width=0.50\textwidth]{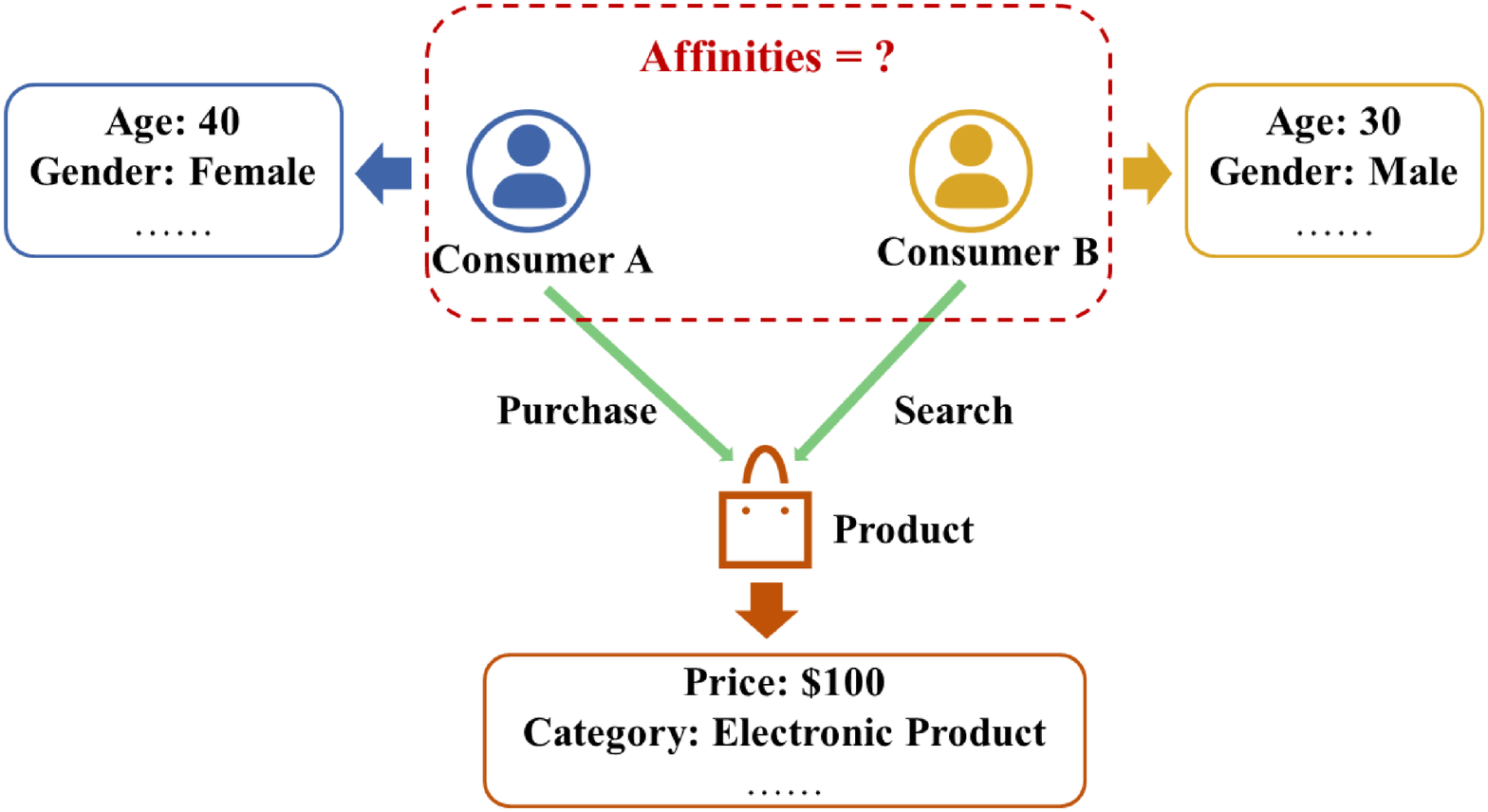}\\
  \caption{An example of a customer-product association graph in a relational database. The affinities between pairs (customers) can be computed from statistics of a random walk based on the cosine correlation of two states on the entire graph. }\label{association}
\end{figure}

Fig.~\ref{association} shows a fragment of the association graph.
The authors study the expected behavior of random walks on the association graph and propose a novel measure of similarity based on the cosine correlation of two states in a random walk. One significant advantage of random walks view is that it can incorporate large amounts of contextual information. Compared with original measures by cross validation experiments, the authors prove that the new measure is more predictive and robust to perturbations.

Fouss et al.~\cite{fouss2005novel,fouss2007random} also use random walks in the movie collaborative recommendation. The authors exploit the graph structure of the relational database to calculate dissimilarity between elements in sets. They compare ten different scoring algorithms. Five of them are based on random walks: the average commute time (CT, normal and PCA-based), the average first-passage time (one-way and return), and the
pseudoinverse of the Laplacian matrix ($\textbf{L}^{+}$).

They introduce a general procedure for computing similarity between elements of a relational database. The authors use movie recommendation as an example to show that ($\textbf{L}^{+}$) almost always provide the best results in comparison with standard methods.

Yildirim et al.~\cite{yildirim2008random} propose a novel item-oriented algorithm called Random Walk Recommender. It is the first time to infer the transition probabilities between items based on their similarity. They first construct an item graph which captures the similarity of items between each other. Then they computes the rank values of items for each user by simulating a random walk on the graph. The rank values can be regarded as ratings between users and items. They prove their method performs significantly better than top-N algorithm~\cite{deshpande2004item} especially when the training data is sparse.

One of the biggest problems in collaborative filtering is the cold-start problem presented by Resnick~\cite{resnick1994grouplens}. It means that it is hard to do collaborative filtering for users who have rated only a very small number of items. Although there are some trust-based methods~\cite{golbeck2005computing,andersen2008trust} trying to solve the problem, the precision is not good enough.

Jamaliz et al.~\cite{jamali2009trustwalker} propose a model called TrustWalker to solve this problem. They combine the trust-based and the item-based collaborative filtering approach to recommendation. It considers not only ratings of the target item, but also those of similar items.

Therefore, users in the trust network will keep a strong trust with the source user and we will get enough ratings at the same time, which will improve the precision of recommendations.

\subsection{Recommender System}
Recommender system is a subclass of information filtering system which attempts to predict users' ratings or preferences for items. It usually uses three ways to produce a list of recommendations: collaborative filtering, content-based filtering, and hybrid filtering.

Gori et al.~\cite{gori2007itemrank} propose ItemRank, which is a scoring algorithm based on random walks. It can be used to rank products according to expected user preferences. They construct a correlation graph of movies. With the help of correlation graph, they can spread users' preferences. This procedure is similar to PageRank. Thus, it can be regarded as a biased version of PageRank designed to be applied to a recommender system.

Gori et al.~\cite{gori2006research} propose PaperRank algorithm based on random walks to solve the paper recommendation problem. Its structure is similar to ItemRank~\cite{gori2007itemrank}. They utilize the model expressed by the citation graph and find out valuable papers related to research topics for researchers. Experiments on the ACM Portal Digital Library dataset demonstrate the outstanding performance of PaperRank.

Xia et al.~\cite{xia2016scientific} propose a method called CARE which incorporates author relations and historical preferences for scientific article recommendation. They assume that some researchers prefer to search articles published by the same authors to find articles they are interested in. The authors build a graph based on the information of co-authors' relationship. Then they employ the random walk with restart to generate a recommendation list. Compared with some baseline algorithms, the algorithm performs better in precision, F1-score, and recall.

Scholar collaboration is very important in academic research, but it is time-consuming to find a valuable collaborator. Xia et al.~\cite{xia2014mvcwalker} propose the MVCWalker method based on random walks to find the most valuable collaborators. The authors use three academic factors to define link importance in academic social networks. Then they perform random walk with restart on the network to get the recommendation list of most valuable collaborators.

\subsection{Link Prediction}
Link prediction in a network refers to how to predict the possibility of links between two nodes in a network that have not yet been connected by network information. Lots of methods have been proposed to solve this problem~\cite{lu2011link,al2011survey}. Liben-Nowell et al.~\cite{liben2007link} compare different methods in link prediction in detail, including hitting time, PageRank, and other variants of random walks.

\par

\par
The computation of hitting time and commute time is time consuming.
To address this problem, Sarkar et al.~\cite{sarkar2012tractable} propose a truncated variant of commute time in the link prediction task. It utilizes the local structure of graphs. Then they propose an algorithm called GRANCH to find out which two nodes will have an edge in the near future. Experiments prove that GRANCH reduces the computation and storage while retaining the performance of methods.

Similarly, Liu et al.~\cite{liu2010link} propose two similarity indices for link prediction based on local random walk: the Local Random Walk index and the Superposed Random Walk index. While maintaining good prediction accuracy, they have lower time complexity.

Backstrom et al.~\cite{backstrom2011supervised} propose supervised random walks. It is a supervised learning task and ranks the nodes based on the network information including rich node and edge attributes. Its purpose is to learn the parameters of the function that assigns the strength of the edge such that a random walker is more likely to reach nodes to which new links will be created in future.

Link prediction also helps researchers find out the potential relation between miRNAs and diseases~\cite{liu2017inferring}. The authors consider the miRNA-Disease heterogeneous network as two overlapping sub-networks: miRNA similarity sub-network and diseases similarity sub-network. They employ random walk with restart to predict miRNA candidates that could potentially be associated with diseases. Cross validation and case analysis show that the method has good prediction performance.

\subsection{Computer Vision}
Computer vision is an interdisciplinary field that deals with how computers can be made to gain high-level understanding from digital images or videos. Its tasks include methods for acquiring, processing, analyzing, and understanding digital images, and extraction of high-dimensional data from the real world.

Meila et al.~\cite{meila2001random} present an approach of image clustering and segmentation based on random walks. The authors focus on pairwise (or similarity-based) clustering and image segmentation. They regard the pairwise similarities as edge flows in a Markov random walk and study the properties of the eigenvectors and values of the transition matrix.

Gorelick et al.~\cite{gorelick2006shape} characterize the shape of a picture using random walks. For each internal pixel, they calculate the value reflecting the mean time required for a random walker beginning at the pixel to reach the boundary. With the help of these calculated values, they extract many properties of the silhouette, such as its part structure, rough skeleton, local orientation, convex part, and concave part.

Grady et al.~\cite{grady2006random,grady2004multi} propose a new algorithm for performing multi-label and interactive image segmentation. The interactive image segmentation means that the user has to label some pixels in the image manually. The algorithm can calculate the probability of the random walker which starts from an unlabeled pixel reaching the pre-labeled pixels. Therefore, a good image segmentation arises from the labels of all the pixels by assigning each pixel to a label with the maximum probability. But the algorithm has some problems, of which is that it requires user-specified seeds. To solve the above problem, Grady~\cite{grady2005multilabel} proposes that combining a prior model into the energy minimization yields an extended random walkers algorithm. It can locate disconnected objects without user-specified labels.

Qiu et al.~\cite{qiu2005image} exploit the properties of the commute time to develop a image clustering and segmentation method. By using the discrete Green's function of graphs, they analyze the cuts of the image from commute time. Qiu et al.~\cite{qiu2006robust} also use commute time to motion track. The main purpose of using commute time as proximity measure is to alleviate the effect of noise on the shape interaction matrix. Commute time is a more robust measure than raw proximity matrix when facing the noise on the shape interaction matrix. They calculate the commute time using the Laplacian eigensystem.

Shen et al.~\cite{shen2014lazy} propose a new image superpixel segmentation approach using LRW algorithm. The authors initialize the seed positions and run the LRW algorithm on the input image to obtain the probabilities of each pixel. Then the boundaries of initial superpixels are obtained with the help of probabilities and the commute time. The new algorithm can segment the weak boundaries and complicated texture regions very well.

Dong et al.~\cite{dong2016sub} present a novel framework based on the subMarkov random walk for interactive seeded image segmentation. It can be regarded as a traditional random walker with some new auxiliary nodes, that makes the framework more flexible. Under this framework, the authors design a new subRW algorithm with label prior to solve the segmentation problem of objects with thin and elongated parts.

Li et al.~\cite{li2016visual} propose a visual tracking algorithm based on random walks on two graph models. Nodes and edges in the graph denote superpixels and the relationships between superpixels, respectively. They incorporate the structural information between target parts and similarity measurements into a structural model to improve the tracking accuracy. It is the first time that visual tracking is treated as Markov random walks~\cite{li2016visual}.

\subsection{Semi-supervised Learning}
Semi-supervised learning is a class of machine learning tasks and techniques. It uses a small amount of labeled data and a large amount of unlabeled data for training. Due to the less human effort and high precision, it is meaningful both in theory and in practice~\cite{zhu2006semi}.

Zhu et al.~\cite{zhu2003semi} present a new approach of semi-supervised learning based on random walks. They do classification task in continuous state space rather than in the discrete label set. The intuition of the approach is that the data points should be labeled as same as their neighbors. The authors' strategy is to employ a real-valued function $f: V\rightarrow R$ on graph $G$ and then to assign labels based on $f$. The function $f$ provides a consistent probabilistic semantics. It is the basis of this semi-supervised classification method. The promising result has shown that the approach can improve the accuracy of classification by exploiting the structure of unlabeled data.

Szummer et al.~\cite{szummer2002partially} find that the partially labeled data may be in the sub-manifold space. The authors hope the measure can incorporate the structure of manifold and the density. Based on these considerations, they present a Markov random walk model to classify the data. The research~\cite{tishby2001data} shows how to change the distance matrix into a Markov process and helps a lot with the construction of graph.

They classify node $j$ with the label $c$ when $c$ maximizes the following formula:
\begin{equation}
c_k=argmax_c \ \sum_{i}P(c\vert i)P_{0\vert t}(i\vert j)
\end{equation}
where $P_{0\vert t}(i\vert j)$ is the probability of a random walk from node $i$ to node $k$, $P(c|i)$ can be estimated by two techniques: maximum likelihood with Expectation Maximization (EM) and maximum margin subject to constraints.

The parameter $t$ in this approach is also important. It denotes the number of transitions which determines the smoothness of a random walk.

However, the choice of $t$ can be tricky and subjective. To overcome this problem, Azran \cite{azran2007rendezvous} presents the rendezvous algorithm.

The author also represents the data points as nodes of a graph and employ the random walk view to do classification.

Different from the work of Szummer et al.~\cite{szummer2002partially}, the labeled points in rendezvous algorithm don't propagate, but absorb the states of the random walk. The probability that each unlabeled data is absorbed by different labeled points can be used to derive the distribution as the transition steps increase to infinity.

Hence, the rendezvous algorithm doesn't bother to choose a good value of the parameter $t$.

\subsection{Network Embedding}
Network embedding can encode nodes or edges to lower dimensional vector representations and keep network structure~\cite{cui2018survey}. It is a promising direction for network representation and can be used to improve performance for downstream tasks.

Inspired by Word2Vec~\cite{mikolov2013efficient}, Perozzi et al.~\cite{perozzi2014deepwalk} propose a new approach called DeepWalk for learning latent vector representations of nodes in a network. DeepWalk uses truncated random walks to extract local information of nodes. Analogy with language models, the sequence of nodes resulted from random walks can be regarded as sentences and the nodes in the network is equal to the words in vocabulary. 

Perozzi also extend \textit{SkipGram} and \textit{Hierarchical Softmax} from Word2Vec to DeepWalk for reducing computation and speeding up convergence rate.

Grover et al.~\cite{grover2016node2vec} find that current feature learning methods cannot adequately express the diversity of connection patterns in the network. Thus they propose Node2Vec which is a novel algorithmic framework for learning feature representations of nodes. It presents a flexible neighborhood sampling strategy based on random walks. In previous methods, considering a random walk that just walks from node $v_i$ to node $v_j$, the single step transition probability of a random walk from node $v_j$ to node $v_k$ is based on the weight $w_{jk}$ of edge $(j, k)$. But node2Vec denotes the unnormalized transition probability $p_{jk}$ as $p_{jk}=\alpha_{pq}(i,k)\dot{w_{jk}}$, and
\begin{equation}
	\alpha_{pq}(i,k)=
	\begin{cases}
	\frac{1}{p} & if \ d_{ik}=0 \\
	1 & if \ d_{ik}=1	\\
	\frac{1}{q} & if \ d_{ik}=2	\\
	\end{cases}
\end{equation}
where $d_{ik}$ is the length of shortest path between nodes $v_i$ and $v_k$. $p$ is the return parameter which controls the likelihood of revisiting a node in the walk. $q$ is the in-out parameter~\cite{grover2016node2vec}.

Actually, the definition of $\alpha_{pq}$ can be regarded as a tradeoff between \textit{breadth-first sampling} (BFS) and \textit{depth-first sampling} (DFS)~\cite{grover2016node2vec}.

\subsection{Element Distinctness}
Element distinctness problem is to tell whether all the elements in a given sequence are distinct. More precisely, it can be described as ``given a series of numbers $x_1,x_2,...,x_N \in [M]$, are there $x_i, x_j \in M$ and $i \neq j$ such that $x_{i}=x_{j}$~\cite{buhrman2001quantum}?"
There is a simple classical algorithm to solve this problem with $Nlog(N)+O(N)$ comparisons. Buhrman et al.~\cite{buhrman2001quantum} present a quantum algorithm to speedup. Their algorithm gives an upper bound of computation cost $O(N^{3/4}log(N))$.

Ambainis~\cite{ambainis2007quantum} improves the quantum way to solve element distinctness with $O(N^{2/3})$ comparisons.The intuition of this optimal quantum algorithm is to construct a graph, and transform the element distinctness problem of finding a marked vertex in the graph. In order to search marked vertex efficiently, the author improves the Grover's quantum search algorithm~\cite{grover1996fast,brassard2002quantum}. The author reuses the information that queries before, and search a marked vertex with $O(N^{2/3})$ comparisons instead of $O(N)$ comparisons in Grover's search algorithm. For the extensions of this algorithms, the authors propose that if we want to find $k$ numbers which are equal in $x_1,x_2,...,x_N$, we can get a quantum walk based algorithms with $O(N^{k/(k+1)})$ queries.

\section{Open Issues}\label{section5}
In this part, we will introduce some major problems of random walks. Most of them are caused by the growing real-world networks.

\subsection{Speed of Random Walk Algorithms}
The time complexity of random walk graph kernel is at least $O(n^{3})$ or $O(m^{2})$ for graph with $n$ nodes and $m$ edges~\cite{kang2012fast}. In an artificially generated graph, this time complexity is acceptable. But it is a disaster on a real-world network since the number of nodes and edges is huge. It is also a challenge for random walk models of which the time complexity are at least $O(n^{2})$. Researchers are already dealing with the problem. Kang et al.~\cite{kang2012fast} propose ARK graph kernels with time complexity $O(n^{2})$ or $O(m)$. There is a prerequisite for this graph kernel. The graph must have lower intrinsic ranks than the order of the graph.

Tong et al.~\cite{tong2006fast} also realize the speed problem in random walk with restart. The random walk with restart algorithm is slow in query time or prohibitive on storage space.

The authors exploit the block-wise community-like structure and the linear correlations of the adjacency matrix of real-world networks. With these two properties, the authors devise B\_LIN to make the random walk with restart faster. This approach not only saves a lot of storage space and computing time, but also preserves good performance.

As we can see, the main idea to cope up with speed of random walk algorithms is to obtain approximate computation instead of accurate computation. We still require more accurate approximate algorithms for random walks.
\subsection{Problem of Main-Memory Volume}
All the fast random walk graph kernels or algorithms are under the consumption that the whole graph can be fit in the main-memory. But with the rapid growth in the scale of the network, this condition can't be satisfied any more. One of the solutions is to divide the graph into several clusters.

There are studies providing some approaches for graph partition and clustering on giant network~\cite{karypis1998fast,karypis1998software}. One of the most popular method is METIS~\cite{karypis1998software}. Since more and more researchers pay attention to the giant network problem, there is a more effective clustering algorithm for graph clustering and a better method to apply random walks on giant network with external memory~\cite{sarkar2010fast}. The author calls the clustering method RWDISK. RWDISK has been proved to be a better way for graph partition on several famous datasets such as Digital Bibliography \& Library Project (DBLP), Citeseer. But these methods still have an unacceptable time latency with respect to enormous graph. There are two ways to solve this issue, partition and using external memory.

\subsection{Computation of Hitting and Commute Time}
As we have mentioned, proximity measures play an important role in network analysis and beyond. The complexity of computing commute time is $O(n^{3})$ which is prohibitive in large real-world graphs. There are some approximations of commute time to reduce the complexity~\cite{brand2005random,sarkar2012tractable}. But we should be careful about these approximate approaches. They can't represent the structure of large real-world graphs or show the connectivity of nodes in large graphs.

Luxburg et al.~\cite{von2010hitting} have shown that commute time can be approximated by simple formula with high accuracy when random geometric graphs (k-nearest neighbor graphs, $\epsilon$-graphs, and Gaussian similarity graph) are large enough. More specifically, commute time $H_{uv}$ can be represented by $1/d_{u}+1/d_{v}$ in large graphs where $d_{u}$ and $d_{v}$ denote the degree of vertex $u$ and vertex $v$, respectively. Thus, the approximations only consider the local density of two nodes rather than the structure information of the whole graph. The authors give two strategies to prove the result: one based on the flow argument of the electric network, and the other based on spectral argument. Both of them prove that approximations of commute time don't take into account any global properties of large graphs. In that case, the effectiveness of approximated commute time is doubtful. The computation of commute time in large graph is still a challenge.

\section{Conclusion}\label{section6}
In this paper, we have presented an overview of random walks from the perspective of computer science, including classical random walks and quantum walks. We first introduce the basic knowledge and some algorithms of classical random walks and quantum walks in a comprehensible way. The typical random walk algorithms are PageRank and its variants. RWR and LRW are also reviewed. They are time-consuming when applied to large real-world graphs. Some methods are developed to accelerate convergence, such as Quadratic Extrapolation, B\_LIN, K-dash. Then two types of algorithms based on quantum walks are discussed: continuous quantum walk based algorithms and discrete quantum walk based algorithms. We make comparisons between classical random walks and quantum walks and find that with the development of quantum computation, the quantum view of random walks accelerates the computation of random walk algorithms significantly.

Random walks can be used to calculate the proximity between two nodes and extract the network topology. It has been proved that random walks play an important role in many scenarios. We explore the applications of random walks in the field of computer science including collaborative filtering, computer vision, network embedding, and so on. Many problems with existing random walk based algorithms are caused by giant networks, such as slow convergence speed, insufficient storage capacity. Further research in this field would be of great help in theoretical and practical application of random walks.

\ifCLASSOPTIONcaptionsoff
  \newpage
\fi



%
\bibliographystyle{IEEEtran}
\bibliography{IEEEabrv,refer}

%

\begin{IEEEbiography}[{\includegraphics[width=1in,height=1.25in,clip,keepaspectratio]{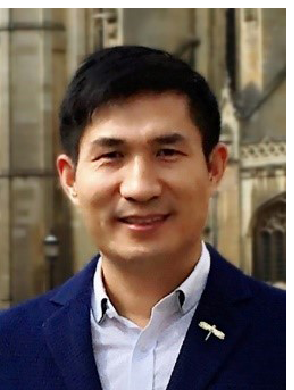}}]{Feng Xia}
(M'07-SM'12) received the BSc and PhD degrees from Zhejiang University, Hangzhou, China. He is currently an Associate Professor of Data Science at Federation University Australia, and on leave from Dalian University of Technology, China, where he is a Full Professor. Dr. Xia has published 2 books and over 200 scientific papers in international journals and conferences. His research interests include data science, big data, knowledge engineering, social computing, and systems engineering. He is a Senior Member of IEEE and ACM.
\end{IEEEbiography}

\begin{IEEEbiography}[{\includegraphics[width=1in,height=1.25in,clip,keepaspectratio]{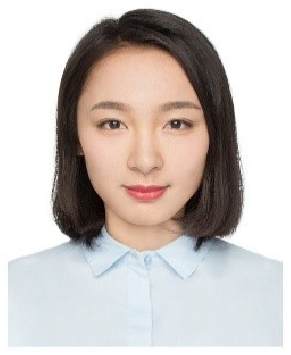}}]{Jiaying Liu}
received the BSc degree in software engineering from Dalian University of Technology, China, in 2016. She is currently working toward the PhD degree in the School of Software, Dalian University of Technology, China. Her research interests include data science, big scholarly data, and social network analysis.
\end{IEEEbiography}

\begin{IEEEbiography}[{\includegraphics[width=1in,height=1.25in,clip,keepaspectratio]{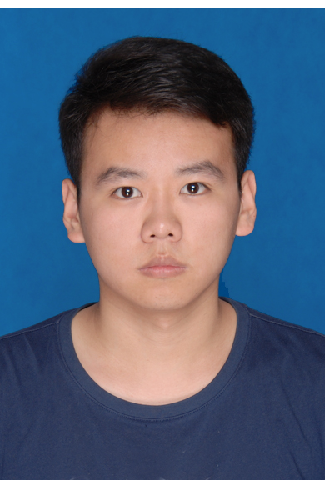}}]{Hansong Nie}
received the BSc degree in electronic information engineering from Dalian Maritime University, China, in 2018. He is currently pursuing the master's degree in in the School of Software, Dalian University of Technology, China. His research interests include big scholarly data, social network analysis, and science of success.
\end{IEEEbiography}

\begin{IEEEbiography}[{\includegraphics[width=1in,height=1.25in,clip,keepaspectratio]{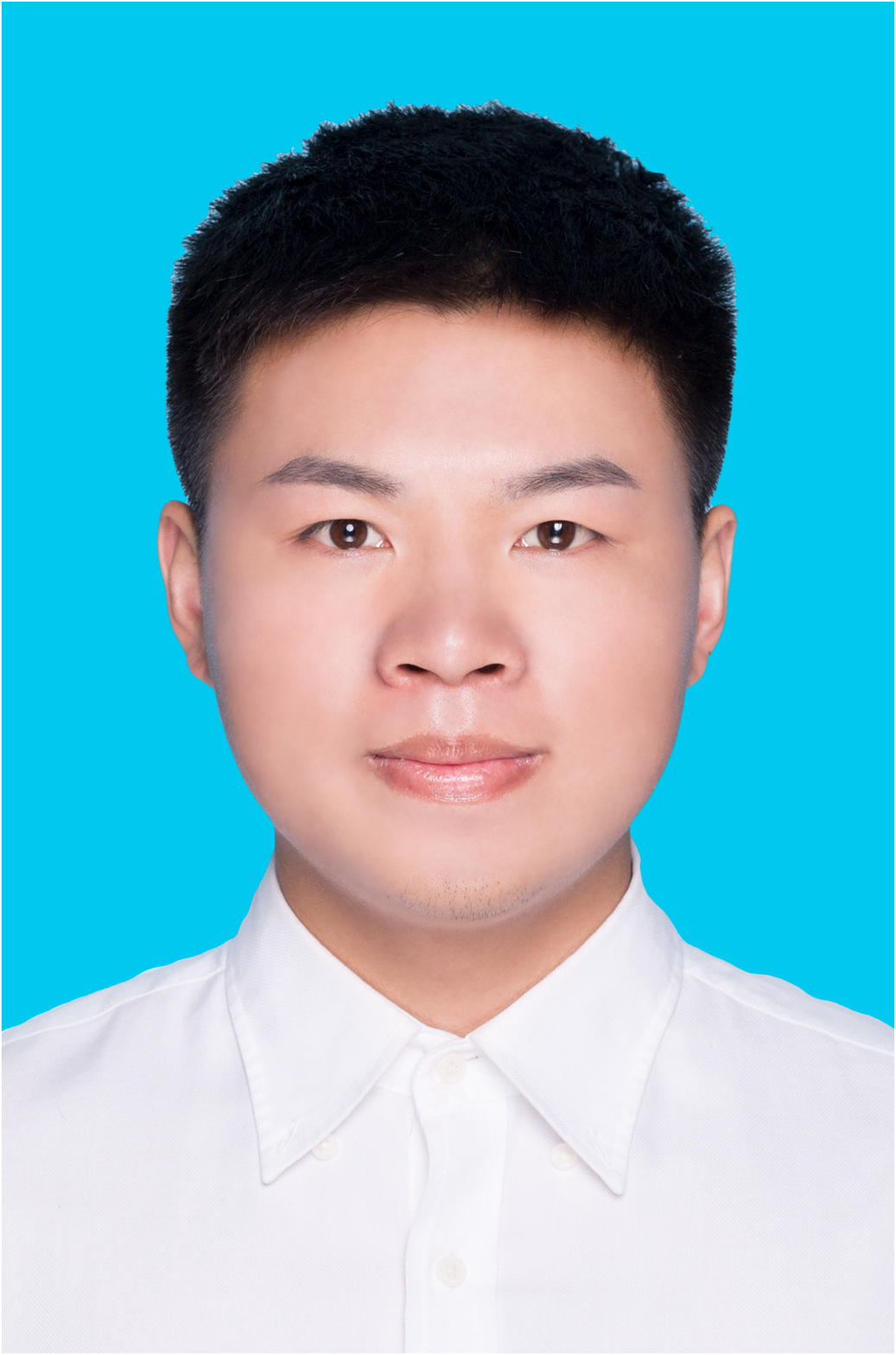}}]{Yonghao Fu}
is currently study as a senior student in Software School of Dalian University of Technology, China. His research interests are Data Mining and Natural Language Processing, and will pursue the master's degree in this field.
\end{IEEEbiography}

\begin{IEEEbiography}[{\includegraphics[width=1in,height=1.25in,clip,keepaspectratio]{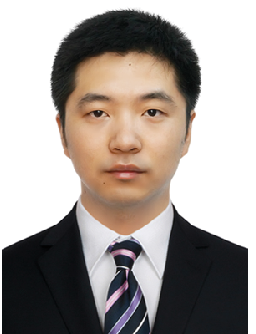}}]{Liangtian Wan}(M'15)
received the BSc degree and the PhD degree in the College of Information and Communication Engineering from Harbin Engineering University, Harbin, China, in 2011 and 2015, respectively. From Oct. 2015 to Apr. 2017, he has been a Research Fellow of School of Electrical and Electrical Engineering, Nanyang Technological University, Singapore. He is currently an Associate Professor of School of Software, Dalian University of Technology, China. He is the author of over 40 papers published in related international conference proceedings and journals. Dr. Wan has been serving as an Associate Editor for IEEE Access and Journal of Information Processing Systems. His current research interests include social network analysis and mining, big data, array signal processing, wireless sensor networks, compressive sensing.
\end{IEEEbiography}

\begin{IEEEbiography}[{\includegraphics[width=1in,height=1.25in,clip,keepaspectratio]{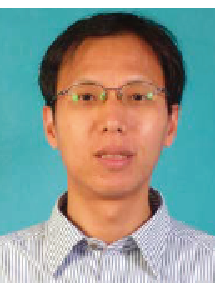}}]{Xiangjie Kong}
(M'13-SM'17) received the BSc and PhD degrees from Zhejiang University, Hangzhou, China. He is currently an Associate Professor in School of Software, Dalian University of Technology, China. He has served as (Guest) Editor of several international journals, Workshop Chair or PC Member of a number of conferences. Dr. Kong has published over 100 scientific papers in international journals and conferences (with 70+ indexed by ISI SCIE). His research interests include computational social sicence, data science, and mobile computing. He is a Senior Member of IEEE and CCF, and a Member of ACM.
\end{IEEEbiography}





\end{document}